\documentclass[reqno,12pt]{amsart} 

\date{\today}

\usepackage{amsmath}
\usepackage{amsfonts}
\usepackage{amssymb}
\usepackage{amsthm}
\usepackage{amstext}
\usepackage{enumerate}
\usepackage{url}
\usepackage{amsbsy}

\newcommand{\bx}{\mathbf{x}} 
\newcommand{\by}{\mathbf{y}}

\newcommand{\rz}{\mathbb{R}}
\newcommand{\N}{\mathbb{N}}
\newcommand{\C}{\mathbb{C}}
\newcommand{\R}{\mathbb{R}}

\theoremstyle{plain}
\newtheorem{thm}{Theorem}[section]{\bf}{\it}
\newtheorem{prop}[thm]{Proposition}{\bf}{\it}
\newtheorem{lemma}[thm]{Lemma}{\bf}{\it}
{\bf}{\it}
\newtheorem{cor}[thm]{Corollary}{\bf}{\it}
\theoremstyle{definition}
\newtheorem*{acknowledgement}{Acknowledgement} 
\newtheorem{remark}[thm]{Remark}{\it}{\rm}
{\bf}{\rm}
{\rm}{\rm}

\newenvironment{pf*}[1]{\par\medskip\noindent\textit{#1}\,:}{\hspace*{\fill}\qed\medskip\par\noindent}   

\title[Analytic structure of solutions to
MC-equations]{Analytic structure of solutions to
  multiconfiguration equations}

\author[S. Fournais, M. and T. Hoffmann-Ostenhof, and T. \O. S\o rensen]
{S. Fournais \and M. Hoffmann-Ostenhof \and T. Hoffmann-Ostenhof \and
T. \O stergaard S\o rensen}

\address[S. Fournais]{Department of Mathematical Sciences, University 
  of Aarhus, Ny Munkegade, Building
  1530, DK-8000 \AA rhus C, Denmark.}
\email{fournais@imf.au.dk}           
\address[S. Fournais on leave from]{CNRS and Laboratoire de
  Math\'{e}matiques d'Orsay, 
Univ Paris-Sud, Orsay CEDEX, F-91405, France.}

\address[M. Hoffmann-Ostenhof]
        {Fakult\"at f\"ur Mathematik,
         Universit\"at Wien,         
         Nordbergstra\ss e 15,
         A-1090 Vienna,
         Austria.} 
\email{maria.hoffmann-ostenhof@univie.ac.at}

\address[T. Hoffmann-Ostenhof]{Institut f\"ur Theoretische
Chemie, W\"ahringer\-strasse 17,
           Universit\"at Wien,
           A-1090 Vienna,
           Austria.}

\address[T. Hoffmann-Ostenhof, 2nd address]{
        The Erwin Schr\"{o}dinger International Institute for 
        Mathematical Physics,
              Boltzmanngasse 9,
              A-1090 Vienna, Austria.}
\email{thoffman@esi.ac.at}

\address[T. \O stergaard S\o rensen]
{Department of Mathematical Sciences,
           Aalborg University,
           Fredrik Bajers Vej 7G,
           DK-9220 Aalborg East, Denmark.}
\email{sorensen@math.aau.dk}

\thanks{\copyright\ 2008 by the
       authors. This article may be reproduced, in its entirety, for
       non-commercial purposes.}

\begin{document}

 \begin{abstract}  
  We study the regularity at the positions of the (fixed) nuclei of
  solutions to (non-relativistic) 
  multiconfiguration equations (including Hartree--Fock) of Coulomb
  systems. We prove the following: Let
  \(\{\varphi_1,\ldots,\varphi_M\}\) be any solution to the
  rank--\(M\) multiconfiguration equations for a molecule with \(L\)
  fixed nuclei at 
  \(R_1,\ldots,R_L\in\R^3\). Then, for any
  \(j\in\{1,\ldots,M\}\), \(k\in\{1,\ldots,L\}\), 
  there exists a neighbourhood \(U_{j,k}\subseteq\R^3\) of \(R_k\),
  and functions \(\varphi^{(1)}_{j,k}, \varphi^{(2)}_{j,k}\), real
  analytic in \(U_{j,k}\), such that
  \[\varphi_j(\bx)=\varphi^{(1)}_{j,k}(\bx)
  +|\bx-R_k|\varphi^{(2)}_{j,k}(\bx)\,,\quad {\bf x}\in U_{j,k}\,.\]  
  A similar result holds for the corresponding electron density.
  The
  proof uses the Kustaan\-heimo--Stiefel transformation, as applied in 
  \cite{KS} to the study of the eigenfunctions of the Schr\"odinger operator
  of atoms and molecules near two-particle coalescence points. 
 \end{abstract}

\maketitle

\section{Introduction and results}
We consider the Hamiltonian of a molecule with $N$ non-relativistic
electrons and \(L\) (static) nuclei of (positive)
  charges \(Z_1,\ldots,Z_L\), fixed at \(R_1,\ldots,R_L\in\rz^3\),
  given by
\begin{align}\label{Hamiltonian}
  H=H(N,Z)&=\sum_{j=1}^{N}\Big\{-\Delta_j
  +V(\bx_j)\Big\}
  +\sum_{1\leq i<j\leq N}\frac{1}{|\bx_{i}-\bx_{j}|}\,,\\
  \label{eq:V}
  V(\bx)&={}-\sum_{k=1}^{L}\frac{Z_k}{|\bx-R_k|}\,.
\end{align}
Here, \(\bx_{j}\in\R^3\) is the coordinate of the \(j\)'th electron
and \(\Delta_j\) is the Laplacian with respect to \(\bx_j\). 
The operator $H$ acts on a dense subspace of the
$N$-particle Hilbert space $\mathcal{H}_{F}=\bigwedge 
_{i=1}^{N}L^{2}(\mathbb{R}^{3};\mathbb{C}^{q})$ of antisymmetric
functions, where $q$ is the number of 
spin states. More precisely, its operator domain is
$\mathcal{D}(H)=\bigwedge  
_{i=1}^{N}W^{2,2}(\mathbb{R}^{3};\mathbb{C}^{q})$ and its quadratic form
domain is $\mathcal{Q}(H)=\bigwedge 
_{i=1}^{N}W^{1,2}(\mathbb{R}^{3};\mathbb{C}^{q})$
\cite{KatoBook,RS2}. 
Since spin is irrelevant for the discussion in
this paper, we let \(q=1\) from now on to simplify
notation. In the case most relevant for physics, namely electrons in a
molecule, $q$ takes the value $2$.

Let \(\mathfrak{q}\) be the quadratic form defined by \(H\), that 
is, for \(\Psi\in\mathcal{D}(H)\),
\(\mathfrak{q}(\Psi,\Psi)=\langle\Psi,H\Psi\rangle\). Then, for
\(\Psi\in\mathcal{Q}(H)\), 
(with
\(\mathbf{X}=(\bx_1,\ldots,\bx_N)\in\R^{3N}\)), 
\begin{align}\label{eq:QM-functional}
 \mathfrak{q}(\Psi,\Psi)
  =\sum_{j=1}^{N}\int_{\R^{3N}}|\nabla_j\Psi(\mathbf{X})|^2\,d\mathbf{X}
   \\&\ +\int_{\R^{3N}}\Big\{\sum_{j=1}^{N}
   V(\bx_j)+\sum_{1\le
       i<j\le N}\frac{1}{|\bx_i-\bx_j|}\Big\}
  \,|\Psi(\mathbf{X})|^2\,d\mathbf{X}\,.
  \nonumber
\end{align}
Here, \(\nabla_j\) is the gradient with respect to \(\bx_j\), and
\(\langle\ 
,\ \rangle\) is the scalar product in \(\mathcal{H}_{F}\subset
L^2(\rz^{3N})\).  
The quadratic form \(\mathfrak{q}\) is bounded from below.
The \textit{quantum ground state energy} is the infimum of this
quadratic form: 
\begin{align}\label{eq:min-QM}  \nonumber
  E^{{\rm QM}}(N,Z):&=\inf \sigma _{\mathcal{H}_{F}}(H)
  \\&=\inf\{\,\mathfrak{q}(\Psi,\Psi)\,|\, \Psi\in \mathcal{Q}(H),
  \langle\Psi,\Psi\rangle=1\} \,.
\end{align}

The Euler--Lagrange equation for the minimization problem
\eqref{eq:min-QM} is nothing but the (stationary) Schr\"odinger
equation, 
\begin{align}
  \label{eq:Schr}
  H\Psi=E\psi\ , \quad\Psi \in \mathcal{D}(H)\,,
\end{align}
with \(E\equiv E^{{\rm QM}}(N,Z)\). A {\it ground state} of the atom is a
solution to \eqref{eq:Schr} for \(E=E^{{\rm QM}}(N,Z)\); {\it excited
states} of the atom are solutions to \eqref{eq:Schr} with \(E>E^{{\rm
    QM}}(N,Z)\). Zhislin \cite{Zhislin} proved the existence of
both ground states and (infinitely many) excited states, when the total
charge \(Z=\sum_{k=1}^{L} Z_k\) satisfies \(N<Z+1\)
(see also~\cite{Friesecke2}). In
particular, in this case the infimum in \eqref{eq:min-QM} is
attained, i.e., {\it minimizers} exist. On the other hand, Lieb
\cite{Lieb84a, Lieb84b} proved that {\it if} minimizers exist, then
\(N<2Z+L\).  

Since, in practice (i.e., numerically), solving \eqref{eq:min-QM}, or
\eqref{eq:Schr}, is unfeasible for even relatively small \(N\),
various approximations to the problem \eqref{eq:min-QM} have been
developed; for a comprehensive 
discussion of approximations in quantum
chemistry, and an extensive literature list, we refer to
\cite{LeBris2,LeBrisLions}. 
We will not discuss the problems
\eqref{eq:min-QM}--\eqref{eq:Schr} 
further in this paper, but rather investigate (in the spirit of
\cite{KS}) the solutions 
to the Euler--Lagrange equations for one of the most used
approximations: The multiconfiguration self-consistent field method
(MC-SCF) (including Hartree--Fock theory). We now discuss this in more
detail.  

In the perhaps most well-known approximation, the {\it Hartree-Fock
  approximation}, instead of minimizing the 
functional \(\mathfrak{q}\) in the entire (linear) $N$-particle space
\(\mathcal{H}_{F}\) (or rather, \(\mathcal{Q}(H)\)), one 
restricts to wavefunctions \(\Psi\) which are pure wedge products,
also called {\it Slater determinants}\,: 
\begin{align}\label{slater}
  \Psi (\bx_{1},\dots,\bx_{N})
  =\frac{1}{\sqrt{N!}}\,\det(u_{i}(\bx_{j}))_{i,j=1}^{N}
  \equiv |u_1\ldots u_N\rangle(\bx_1,\ldots,\bx_N)
  \,, 
\end{align}
with $\{u_{i}\}_{i=1}^N\subset
W^{1,2}(\mathbb{R}^{3})$,  orthonormal in
$L^{2}(\mathbb{R}^{3})$ (called {\it orbitals}). Notice 
that this way, \(\Psi\in\mathcal{H}_{F}\) and
$\|\Psi\|_{L^{2}(\mathbb{R}^{3N})}=1$.   

The {\it Hartree--Fock ground state energy} is the infimum of the
quadratic form \(\mathfrak{q}\) defined by \(H\) over such Slater 
determinants:  
\begin{align}\label{eq:HF-energy}
  &E^{{\rm HF}}(N,Z):=
  \inf \{\,\mathfrak{q}(\Psi,\Psi) \,|\, \Psi \in \mathcal{S}_N\}\,,\\
  \label{eq:HF-space}
  &\mathcal{S}_N=\big\{\,\Psi=|u_1\ldots u_N\rangle \,\big|\ 
  u_i\in W^{1,2}(\R^3)\,,\, 
  (u_i,u_j)=\delta_{ij}\,\big\}\,,
\end{align}
where \((\,\cdot\,,\,\cdot\,)\)  is the scalar product in
\(L^2(\rz^{3})\).
Clearly, \(  E^{{\rm HF}}(N,Z)\ge   E^{{\rm QM}}(N,Z)\). In fact,
strict inequality holds \cite{LeBris}. Inserting \(\Psi\) of the 
form in \eqref{slater} into \eqref{eq:QM-functional} yields
\begin{align}
  \label{eq:HF-functional}\nonumber
  \mathcal{E}^{\rm HF}(u_1,\ldots,u_N):&=\mathfrak{q}(\Psi,\Psi) 
  \\&=\sum_{j=1}^{N}\int_{\R^3}\big\{|\nabla
  u_j(\bx)|^2+V(\bx)
  |u_j(\bx)|^2\big\}\,d\bx
  \\&   +\int_{\R^3}\int_{\R^3}
   \frac{\rho(\bx)\rho(\by)}{|\bx-\by|}\,d\bx d\by
  -\int_{\R^3}\int_{\R^3}
   \frac{|\gamma(\bx,\by)|^2}{|\bx-\by|}\,d\bx d\by\,,
   \nonumber
\end{align}
where \(\rho\) is the {\it density} and \(\gamma\) is the {\it density
  matrix} of \(\Psi\), given by
\begin{align}
  \label{eq:rho-gamma}
  \gamma(\bx,\by)=\sum_{i=1}^{N}\overline{u_i(\by)}u_i(\bx)\ , 
  \qquad \rho(\bx)=\gamma(\bx,\bx)=\sum_{i=1}^{N}|u_i(\bx)|^2\,.
\end{align}
 With \(\mathcal{E}^{\rm HF}\) defined this way, the minimization problem
 \eqref{eq:HF-energy}--\eqref{eq:HF-space} can be formulated as 
 \begin{align}\label{eq:HF-min}
   E^{{\rm HF}}(N,Z)
   &=\inf\{\,\mathcal{E}^{\rm HF}(u_1,\ldots,u_N)\,|\, 
  (u_1,\ldots,u_N)\in \mathcal{M}_N\}\,,
  \\
  \mathcal{M}_N&=\big\{\, 
  (u_1,\ldots,u_N)\in [W^{1,2}(\R^3)]^N
  \,\big|\
  (u_i,u_j)=\delta_{ij}\,\big\}\,. \label{eq:HF-constraints}
 \end{align}
Both the energy functional \(\mathcal{E}^{\rm HF}\)
and the space \(\mathcal{M}_N\) are nonlinear, but the orbitals
\(\{u_i\}_{i=1}^{N}\) depend only on \(\bx\in\R^3\), whereas \(\Psi\) in
\eqref{eq:min-QM} depends on \(\mathbf{X}\in\R^{3N}\). It is this
reduction in the dimension of the variables which makes the problem
\eqref{eq:HF-min}--\eqref{eq:HF-constraints} more tractable in pratice
(i.e., numerically) than \eqref{eq:min-QM}.

The existence of minimizers (again, when \(Z>N-1\)) for the problem
\eqref{eq:HF-min}--\eqref{eq:HF-constraints} (these 
are not unique since \(\mathcal{E}^{\rm HF}\) is not convex; see also
below) was first proved by  
Lieb and Simon~\cite{LiebSimonHF}. The Euler--Lagrange equations of
the problem \eqref{eq:HF-min}--\eqref{eq:HF-constraints} are the {\it
  Hartree--Fock equations} (HF--equations), 
\begin{align}\label{eq:HF}
  \big({}-\Delta&+
  V\big)\varphi_i(\bx)
  +\Big(\sum_{j=1}^{N}\int_{\R^3}\frac{|\varphi_j(\by)|^2}{|\bx-\by|}
  \,d\by\Big)\varphi_i(\bx) 
  \\&-
  \sum_{j=1}^{N}
  \Big(\int_{\R^3}
  \frac{\varphi_j(\by)\varphi_i(\by)}{|\bx-\by|}\,d\by\Big)\varphi_j(\bx)
  =\varepsilon_i\varphi_i(\bx)\ , \quad 1\le i\le N\,.
  \nonumber
\end{align}
Here, the \(\varepsilon_i\)'s are the Lagrange multipliers of the
orthonormality constraints in \eqref{eq:HF-constraints}. Note that the
naive Euler--Lagrange equations are more complicated than 
\eqref{eq:HF}, but since both the functional \(\mathcal{E}^{\rm HF}\)
in \eqref{eq:HF-functional}
and the ortogonality constraints in \eqref{eq:HF-constraints} are
invariant under unitary transformations (that is, if \((u_1,\ldots,u_N)\in\mathcal{M}_N\) and
\((\tilde{u}_1,\ldots,\tilde{u}_N)=U(u_1,\ldots,u_N)\) for \(U\) an
\(N\times N\) unitary matrix, then
\(\mathcal{E}^{\rm
  HF}(\tilde{u}_1,\ldots,\tilde{u}_N) = \mathcal{E}^{\rm
  HF}(u_1,\ldots,u_N)\) and \((\tilde{u}_1,\ldots,\tilde{u}_N)\in
\mathcal{M}_N\)), 
the {\it matrix} of Lagrange multiplers
due to \eqref{eq:HF-constraints} may be diagonalised without loss of
generality, which turns the Euler--Lagrange equations into
\eqref{eq:HF}. 

In \cite{LiebSimonHF} it was also proved that if
\((\varphi_1,\ldots,\varphi_N)\in\mathcal{M}_N\) is a
minimizer of the 
problem \eqref{eq:HF-min}--\eqref{eq:HF-constraints}
then \(\{\varphi_1,\ldots,\varphi_N\}\) satisfies
\eqref{eq:HF}; they are called {\it ground state solutions} of
\eqref{eq:HF}.  Lions~\cite{Lions} proved (also for \(Z>N-1\)) the 
existence of {\it saddle points},
namely, an infinite sequence
\(\{\pmb{\varphi}_n\}_{n\in\N}=\{\varphi_1^{n},\ldots,\varphi_N^{n}\}_{n\in\N}\)
of solutions of 
\eqref{eq:HF}.
(We refer to 
\cite{Lewin} for a discussion of the relationship between these saddle
points, and the earlier mentioned excited states.) Note that
\eqref{eq:HF} can be re-formulated as 
\begin{align}
  \label{eq:HF-operator eigen}
  h_{\pmb{\varphi}}\varphi_i  = \varepsilon_i\varphi_i\ , \quad
  1\le i\le N\,,
\end{align}
with \(h_{\pmb{\varphi}}\) the {\it Hartree-Fock operator associated
  to \(\pmb{\varphi}=\{\varphi_1,\ldots,\varphi_N\}\)}, given by
\begin{align}
  \label{eq:HF-operator}
  h_{\pmb{\varphi}}u=\big({}-\Delta
  +V\big)u +
  R_{\pmb{\varphi}}u-K_{\pmb{\varphi}}u\,,
\end{align}
where \(V\) is given by \eqref{eq:V}, \(R_{\pmb{\varphi}}u\) is the
{\it direct interaction}, given by the multiplication operator defined
by   
\begin{equation}\label{Rgamma}
  R_{\pmb{\varphi}}(\bx):=\sum_{j=1}^{N}\int_{\mathbb{R}^3}
  \frac{|\varphi_j(\by)|^2}{|\bx-\by|}\,d
  \by\,
\end{equation}
and \(K_{\pmb{\varphi}}u\) is the {\it exchange term}, given by
the integral operator
\begin{equation}
  (K_{\pmb{\varphi}}u)(\bx)=\sum_{j=1}^{N}\Big(\int_{\mathbb{R}^3}
  \frac{\overline{\varphi_j(\by)}u(\by)}{|\bx-\by|}\,d
  \by\Big)\varphi_{j}(\bx)\,.
\end{equation}
The equations \eqref{eq:HF-operator eigen} are called the {\it
  self-consistent Hartree-Fock equations}. If \(\Psi\) 
is a minimizer for
the problem \eqref{eq:HF-energy}--\eqref{eq:HF-space}, 
then 
\(\Psi\) can be written as \(\Psi=|\varphi_1\ldots\varphi_N\rangle\) 
with the 
\(\varphi_i\)'s solving \eqref{eq:HF-operator eigen}, with
\(\varepsilon_1\le\varepsilon_2\le \cdots\le\varepsilon_N<0\) the \(N\)
lowest eigenvalues of the ope\-rator \(h_{\pmb{\varphi}}\)
\cite{LiebSimonHF}. 
\begin{remark}\label{rem:Hartree}{\rm 
We note that Hartree originally \cite{Hartree} studied the simpler
equations 
\begin{align}
  \label{eq:H}
  \big({}-\Delta+V\big)\varphi_i(\bx)+\Big(\sum_{j\neq
    i}\int_{\R^3}\frac{|\varphi_j(\by)|^2}{|\bx-\by|}
  \,d\by\Big)\varphi_i(\bx) =\varepsilon_i\varphi_i(\bx)\,,\\
  1\le i\le N\,,
  \nonumber
\end{align}
called the  \emph{Hartree equations} (H--equations). He derived these
without going through a
minimization in the variational principle, a refinement which is due
to Slater \cite{Slater}: Ignoring the Pauli principle, \eqref{eq:H} are
the Euler--Lagrange equations for minimizing the functional 
\begin{equation}\label{eq:H-functional}
  \mathcal{E}^{H}(u_1,\ldots,u_N)=\mathfrak{q}(\Psi,\Psi)
\end{equation}
over wavefunctions \(\Psi\) of the form
\begin{equation}\label{eq:product-state}
  \Psi(\bx_1,\ldots,\bx_N)=\prod_{i=1}^{N} u_i(\bx_i)\ , \quad
  u_i\in W^{1,2}(\R^3)\,.
\end{equation}
Fock \cite{Fock} and Slater \cite{Slater} then independently realised how
to introduce the Pauli principle (by using \(\Psi\)'s of the form in
\eqref{slater}), which led to the Hartree--Fock equations in
\eqref{eq:HF}.
} 
\end{remark}
In the {\it multiconfiguration self-consistent field method} (MC-SCF)
one aims to recover more generality on the wavefunction \(\Psi\) by
minimizing \(\mathfrak{q}(\Psi,\Psi)\) in \eqref{eq:QM-functional} on
{\it finite sums} of Slater 
determinants (see \eqref{slater})
instead of only on a single Slater determinant as in
Hartree--Fock theory (of course any \(\Psi\in\mathcal{Q}(H)\) is an
{\it infinite} sum of Slater determinants). More precisely, for \(M\ge
N\), \(M,N\in\N\),  the set of admissible wavefunctions is limited to the
\(\Psi\)'s which are linear combinations of Slater determinants of
length \(N\), built out of \(M\) orbitals. The minimization problem
then becomes
\begin{align}
  \label{eq:MCSCF-functional}
  E^{{\rm MCSCF}}_M&(N,Z)
  =\inf\{\,\mathfrak{q}(\Psi,\Psi)\,|\, \Psi\in \mathcal{S}_N^M\} \,,\\ 
  \label{eq:sum-Slater}
  \mathcal{S}_N^M&=\big\{\,
  \Psi=\!\!\!\!\!\!
  \sum_{I=\{i_1<i_2<\cdots<i_N\}\subset\{1,\ldots,M\}}
  \!\!\!\!\!\!\!\!\!\!\!\!\!\!\!
  c_I |u_{i_1}\ldots u_{i_N}\rangle
  \,  \big|\ 
  u_i\in W^{1,2}(\R^3)\,,\\\,  & \qquad\qquad\qquad\qquad\qquad
 (u_i,u_j)=\delta_{ij}\,,\,
 c_I\in\C\,,\,\sum_{I}|{c_{I}}|^2=1\big\}\,.\nonumber
\end{align}
Note that
\(\mathcal{S}_N=\mathcal{S}_N^N\subset
\mathcal{S}_N^M\), \(M\ge N\) (see \eqref{eq:HF-space}).  
Also, clearly 
\begin{align}\label{eq:three energies}
  E^{{\rm HF}}(N,Z)
  =E^{{\rm MCSCF}}_N(N,Z)\ge E^{{\rm MCSCF}}_M(N,Z)\ge E^{{\rm QM}}(N,Z)\,.
\end{align}
In fact, strict inequality holds also in the last inequality
\cite{Friesecke}. 

One can express the energy \(\mathfrak{q}(\Psi,\Psi)\) for
\(\Psi\in\mathcal{S}_N^M\) as a (nonlinear) functional of the
\(c_I\)'s and the 
\(u_i\)'s (see \cite[(6)]{Lewin}), but since this is somewhat
complicated, and immaterial for our discussion, we shall refrain from
doing so here.

The existence of minimizers (provided \(Z>N-1\)) for the problem
\eqref{eq:MCSCF-functional}--\eqref{eq:sum-Slater}
was proved by Friesecke~\cite{Friesecke2} (and for a related case by
Le Bris~\cite{LeBris}).
The corresponding Euler--Lagrange equations, called the {\it
  multiconfiguration equations} (MC equations),  are
\begin{align}
  \label{eq:MC-SCF1}\nonumber
  &\gamma_i\big({}-\Delta+V\big)\varphi_i 
  +\sum_{j,k,\ell=1}^{M}\Big(A_{ijk\ell}\int_{\R^3}
  \frac{\varphi_k(\by)\overline{\varphi_\ell(\by)}}{|\bx-\by|} 
  \,d\by\Big) \varphi_{j}
  \\&\qquad\qquad\qquad\qquad\qquad\qquad\qquad
  {\ }=\sum_{j=1}^M \lambda_{ij}\varphi_j\,,\quad
  1\le i\le M\,,\\
   &\sum_{J=\{j_1<j_2<\ldots<j_N\}\subset\{1,\ldots,M\}} 
   \!\!\!\!\!\!\!\!\!\!\!\!\!\!\!\!\!\!\!
      H_{IJ}c_J=Ec_I\, ,\
   I=\{i_1<i_2<\cdots<i_N\}\subset\{1,\ldots,M\}\,.\label{eq:MC-SCF2}
\end{align}
The first equation \eqref{eq:MC-SCF1} is a system of \(M\) nonlinear
partial differential 
equations. They are the Euler--Lagrange equations for the
\(\varphi_i\)'s. Here, the coefficients \(\gamma_i>0\) and
\(A_{ijk\ell}\in\C\) are explicit functions of the \(c_I\)'s, and the
\(\lambda_{ij}\)'s are Lagrange multipliers of the orthonormality
constraints on the \(\varphi_i\)'s in \eqref{eq:sum-Slater}. The second
equation \eqref{eq:MC-SCF2} is an eigenvalue problem---the
Euler--Lagrange equations for 
the \(c_I\). Here, the coefficients \(H_{IJ}\) in the equations for
the \(c_I\)'s are explicit functions of the \(\varphi_i\)'s, and \(E\)
is the Lagrange multiplier of the normalisation condition for the
\(c_I\)'s in \eqref{eq:sum-Slater}. The details of this are immaterial
for our discussion; we refer to \cite{Lewin, Friesecke2}. For a derivation of
these equations, see \cite[Appendix 1]{Friesecke2}.

As in the case of the Hartree--Fock equations, the equations
\eqref{eq:MC-SCF1}--\eqref{eq:MC-SCF2} can be written in a more
compact form:
\begin{align}
  \label{eq:MC-compact1}
  \big(({}-\Delta+V)\Gamma+W_\Phi\big)\cdot\Phi&=\Lambda\cdot\Phi\,,
  \\\label{eq:MC-compact2}
  H_{\Phi}\cdot c&=Ec\,,
\end{align}
where \(\Phi=(\varphi_1,\ldots,\varphi_M)^{T}\) and
\(c=(c_I)\in\R^{\binom{M}{N}}\). Here, 
\(\Lambda=(\lambda_{ij})_{1\le i,j\le M}\), and \(\Gamma\) and \(W_{\Phi}\) are
\(M\times M\) matrices (\(\Gamma\) constant, \(W_\Phi\) dependent on
\(\bx\in \R^3\)), given in terms of the \(\gamma_i\)'s and the
\(A_{ijk\ell}\)'s in \eqref{eq:MC-SCF1}--\eqref{eq:MC-SCF2}. Again, we
refer to \cite{Lewin} for more details.

The existence of saddle points, i.e., an infinite sequence
\[\{c_n,\pmb{, \varphi}_n\}_{n\in\N}=\{(c_I)_n;\varphi_1^n,\ldots,\varphi_M^n\}_{n\in\N}\]
of solutions to 
\eqref{eq:MC-SCF1}--\eqref{eq:MC-SCF2} was proved by
Lewin~\cite{Lewin} (again, provided \(Z>N-1\)).

A natural mathematical question is to study the regularity properties
of solutions to the multiconfiguration
equations (including the Hartree--Fock equations). However, this question 
is also of practical interest, since regularity properties of the
solutions have influence on the convergence properties of various
numerical schemes. We refer to \cite{LeBris2,LeBrisLions} for
discussions on this. 

It was proved in \cite[Theorem 3.2]{LiebSimonHF} that if
\(\pmb{\varphi}=\{\varphi_1,\ldots,\varphi_N\}\) is a solution of the
Hartree--Fock equations \eqref{eq:HF-operator eigen}, then the
\(\varphi_i\)'s are globally Lipschitz continuous, i.e.,
\(\varphi_i\in C^{0,1}(\R^3)\). 
This also holds for solutions to the
Hartree equations \cite[Theorem 3.1]{LiebSimonHF} (see also \cite[{\it
  Remarks} 4) p. 192]{LiebSimonHF}). 
The proof readily extends to solutions of the
multiconfiguration equations. Note also that it was proved in
\cite{Lions} (for HF) and in 
\cite{Lewin} (for MC) that the \(\varphi_i\)'s belong to
\(W^{2,p}(\R^3)\) for all \(p\in[2,3)\) and consequently, by 
the Sobolev inequality \cite[Theorem 6 (ii)]{Evans}, to
\(C^{\alpha}(\R^3)\) for all \(\alpha\in(0,1)\).

Furthermore, the \(\varphi_i\)'s are real analytic away from the
positions of the nuclei,  
i.e., \(\varphi_i\in
C^{\omega}(\R^3\setminus\{R_1,\ldots,R_L\})\). This was first proved 
in (the preprint version of) \cite{Lewin}, for 
solutions to the multiconfiguration equations
\eqref{eq:MC-SCF1}--\eqref{eq:MC-SCF2}
(see also \cite{Friesecke});
it was conjectured in \cite{LiebSimonHF}, where smoothness
(\(\varphi_i\in C^{\infty}(\R^3\setminus\{R_1,\ldots,R_L\})\)) was proved. 
Note also that if \(\pmb{\varphi}=\{\varphi_1,\ldots,\varphi_N\}\) is a
solution to \eqref{eq:HF-operator eigen}, and if
\(\varphi\) satisfies \(h_{\pmb{\varphi}}\varphi=\varepsilon\varphi\),
then \(\varphi\) has the same 
regularity properties as those of the \(\varphi_i\)'s discussed above.

The main result of this paper is the following theorem, which completely
settles the regularity properties at the positions \(R_1,\ldots,R_L\)
of the nuclei of all solutions to
the multiconfiguration equations
\eqref{eq:MC-SCF1}--\eqref{eq:MC-SCF2} (including the Hartree--Fock
equations 
\eqref{eq:HF}). We denote by \(B_3(R,r)\subset\R^3\) the ball of
radius \(r>0\) with centre at \(R\in \R^3\).

\begin{thm}\label{thm:main}
Let \(\{(c_I);\varphi_1,\ldots,\varphi_M\}\) be a solution to the
multiconfiguration equations \eqref{eq:MC-SCF1}--\eqref{eq:MC-SCF2}. 

Then, for all \(j\in\{1,\ldots,M\}\) and \(k\in\{1,\ldots,L\}\), there
exist \(\varepsilon\equiv\varepsilon_{j,k}>0\) and real analytic  functions
\(\varphi_{j,k}^{(1)}, \varphi_{j,k}^{(2)}:B_3(R_k,\varepsilon)\to
\C\), that is, \(\varphi_{j,k}^{(1)}, \varphi_{j,k}^{(2)}\in
C^{\omega}(B_3(R_k,\varepsilon))\), such that
\begin{align}
  \label{eq:analytic}
  \varphi_{j}(\bx)=\varphi_{j,k}^{(1)}(\bx)+|\bx-R_k|\varphi_{j,k}^{(2)}(\bx)
  \ , \quad\bx\in B_3(R_k,\varepsilon)\,.
\end{align}
\end{thm}

\ 

\begin{remark}
 \(\, \)
\begin{enumerate}{\rm 
\item[\rm (i)] For simplicity of notation, we have stated everything
  only in the spinless case. It will be obvious that
  the proof of Theorem~\ref{thm:main} also works in the general case
  of spin \(q\). It will also be clear that the result also holds for
  solutions to the Hartree equations \eqref{eq:H}.
\item[\rm (ii)] The result of Theorem~\ref{thm:main} immediately
  implies regularity results for the many-body wavefunction
  \(\Psi\) generated by \((c_I)\) and
  \(\{\varphi_1,\ldots,\varphi_M\}\) (see \eqref{eq:sum-Slater}). 
  For recent results on the regularity properties of the {\it true}
  minimizer \(\Psi\) (i.e., for the problem \eqref{eq:min-QM}) and of excited
  states, we refer to \cite{Monster,KS}. 
  The proof of Theorem~\ref{thm:main} uses the Kustaan\-heimo--Stiefel
  transformation, as applied in \cite{KS} to study these eigenfunctions of
  the Schr\"odinger operator of atoms and molecules (that is,
  solutions to \eqref{eq:Schr}) near two-particle
  coalescence points. 
}
\end{enumerate}
\end{remark}

\begin{remark}{\rm 
Partial results on the {\it asymptotic} regularity at the positions of
the nuclei of solutions to
Hartree--Fock equations were recently given in \cite{Flad2}; more
precisely, estimates of the form
\begin{align}
  \label{eq:asy-est}
  \big|\partial_{\bx}^{\beta}\varphi_j(\bx)\big|\le C_{j,k,\beta,\varepsilon_{j,k}}
  |\bx-R_k|^{1-|\beta|}\,,
\end{align}
for \(|\beta|\ge 1\) and \(\bx\in B_3(R_k,\varepsilon_{j,k})\)
for some \(\varepsilon_{j,k}>0\),
were proved to hold for {\it certain} solutions to the
Hartree--Fock equations, obtained by the so-called
\emph{level-shifting algorithm} \cite{CancesLeBris}. We shall not
discuss 
this in detail here, but just point out that Theorem~\ref{thm:main}
implies that \emph{any} solution to the Hartree--Fock equations (and,
more generally, to the multiconfiguration equations) satisfies the
estimate \eqref{eq:asy-est}. This fact is relevant for the study in
\cite{Flad} of the use of tensor product wavelets in the
approximation of Hartree--Fock eigenfunctions. The result of
Theorem~\ref{thm:main} 
is, however, much stronger than
\eqref{eq:asy-est}. }
\end{remark}

 Theorem~\ref{thm:main} immediately implies similar regularity
 properties for the corresponding (electron) {\it density}. More
 precisely, for \(\Psi\in L^{2}(\R^{3N})\), 
 define \(\rho\equiv\rho_{\Psi}\) by (recall that
 \(\mathbf{X}=(\bx_1,\ldots,\bx_N)\in\R^{3N}\)) 
  \begin{align}
    \label{eq:def-density}
    \rho(\bx)=\sum_{j=1}^{N}\int_{\R^{3N}} |\Psi(\mathbf{X})|^2
     \delta(\bx-\bx_j)\,d\mathbf{X}\,. 
  \end{align}
For \(\Psi\in \bigwedge_{i=1}^{N}L^{2}(\mathbb{R}^{3})\), this becomes
\begin{align}\label{eq:density-anti-sym}
  \rho(\bx)=N\int_{\R^{3N-3}}|\Psi(\bx,\bx_2,\ldots,\bx_N)|^2
  \,d\bx_2\cdots d\bx_N\,.
\end{align}
For \(\Psi\) a Slater determinant, \(\rho\) was given in
\eqref{eq:rho-gamma}; for \(\Psi\) a {\it product state} (see
\eqref{eq:product-state}), \(\rho\) is also given by
\eqref{eq:rho-gamma}, whereas for \(\Psi\) a linear combination of
Slater determinants of 
length \(N\), built out of \(M\) functions (see
\eqref{eq:sum-Slater}), \(\rho\) becomes
\begin{align}\label{eq:density-MC}
  \rho(\bx)=\sum_{j=1}^{M}n_j|\varphi_j(\bx)|^2\,,\quad
  n_j=\sum_{I\ni j}c_I{}^2\,.
\end{align}
Since, for any solution of \eqref{eq:MC-SCF1}--\eqref{eq:MC-SCF2}, the
orbitals are real analytic 
away from the positions of the nuclei, the same holds for the
corresponding density \(\rho\) (i.e., \(\rho\in
C^{\omega}(\R^3\setminus\{R_1,\ldots,R_L\})\)), defined by
\eqref{eq:density-MC} (since these are {\it finite} sums). The
following corollary to Theorem~\ref{thm:main} completely settles the
regularity properties of \(\rho\) at the positions \(R_1,\ldots,R_L\)
of the nuclei. 
\begin{cor}\label{cor:density}
Let \(\{(c_I);\varphi_1,\ldots,\varphi_M\}\) be a solution to the
multiconfiguration equations \eqref{eq:MC-SCF1}--\eqref{eq:MC-SCF2},
and let \(\rho\) be the corresponding electron density, given by
\eqref{eq:density-MC}. 

Then for all \(k\in\{1,\ldots,M\}\) there exist \(\varepsilon_k>0\) 
and real analytic  functions 
  \(\rho_{1}, \rho_{2}:B_3(R_k,\varepsilon_k)\to \R\) 
(i.e., \(\rho_{1}, \rho_{2}\in
C^{\omega}(B_3(R_k,\varepsilon_k))\)), such that
\begin{align}
  \label{eq:density-analytic}
  \rho(\bx)=\rho_{1}(\bx)+|\bx-R_k|\,\rho_{2}(\bx)
  \quad \text{ for all }\ \bx\in B_3(R_k,\varepsilon_k)\,.\\
  \nonumber
\end{align}
\end{cor}

\begin{remark}\label{rem:density}
{\rm Note that the corresponding question for the density \(\rho\)
  (given by \eqref {eq:density-anti-sym}) of the {\it true} minimizer
  of \eqref{eq:min-QM} as well as of excited states---that is,
  solutions to \eqref{eq:Schr}---remains open. In this case, the
  density is known to be real analytic away from the positions of the
  nuclei (i.e., \(\rho\in
  C^{\omega}(\R^3\setminus\{R_1,\ldots,R_L\})\)) (see
  \cite{analytic}), and partial results on the behaviour in the
  vicinity of the nuclei were obtained in \cite{non-iso,thirdder}. 
}
\end{remark}

\section{Proof of the main theorem}
\label{sec:proof}
As mentioned in the introduction the proof of Theorem~\ref{thm:main}
is based on the Ku\-staan\-heimo-Stiefel (KS) transform. We will
'lift' the multiconfiguration equations \eqref{eq:MC-SCF1} to new
coordinates using that 
transform. The solutions to the new equations will be real analytic
functions. By projecting to the original coordinates we get the
structure result in Theorem~\ref{thm:main}. The latter fact was proved in
\cite{KS} (see Proposition~\ref{prop:one particle} below).

The KS--transform \(K:\R^4\to\R^3\) is defined by 
\begin{align}
  \label{eq:KSbis}
  K(\by)=\left(\begin{matrix}
   y_1^2-y_2^2-y_3^2+y_4^2\\
   2(y_1y_2-y_3y_4)\\
   2(y_1y_3+y_2y_4)
   \end{matrix}\right)\ , \quad
   \by=(y_1,y_2,y_3,y_4)\in\R^4\,.
\end{align}
It is a simple computation to verify that
\begin{align}
  \label{eq:norms}
  |K(\by)|:=\|K(\by)\|_{\R^3} = \|\by\|_{\R^4}^2=:|\by|^2\,  \text{
    for all 
  }\by\in\R^4\,. 
\end{align}
Let \(f:\R^3\to\C\) be any
\(C^2\)-function, and define, with \(K\) as above,
\begin{align}\label{eq:def-compo}
  f_K:\R^4\to\C\ , \qquad f_K(\by):=f(K(\by))\,.
\end{align}
Then for all \(\by\in\R^4\setminus\{0\},\) (see \cite[Lemma~3.1]{KS}),
\begin{align} \label{eq:Laplace-formula}
  (\Delta f)(K(\by))&=\frac{1}{4|\by|^2}\,\Delta f_K(\by)\,.
\end{align}

\begin{pf*}{Proof of Theorem~\ref{thm:main}}
We prove the theorem in the case \(k=1\). We assume
without loss of generality (make a linear transformation in \(\R^3\))
that \(R_1\equiv0\in\R^3\). 

Assume \(\{(c_I); \varphi_1,\ldots,\varphi_M\}\) solves the
multiconfiguration equations \eqref{eq:MC-SCF1}--\eqref{eq:MC-SCF2}. 
Define 
\begin{align}\label{def:phi-ij}
  \phi_{k,\ell}:=(\varphi_k\overline{\varphi_\ell})*\frac{1}{|\cdot|}\
  ,\quad 
  k,\ell\in\{1,\ldots,N\}\,.
\end{align}
Then  \eqref{eq:MC-SCF1} can be rewritten
\begin{align}
  \label{eq:HF-augmented1}
  \gamma_i\big({}-\Delta_{\bx}+V\big)\varphi_{i}+\sum_{j,k,\ell=1}^{M} 
  A_{i j k \ell}\,\phi_{k,\ell}\,\varphi_j 
  &= \sum_{j=1}^{M}\lambda_{ij}\varphi_{j}
  \,,\ 
  1\le i \le M\,,\\
  \label{eq:HF-augmented2}
  {}-\Delta_{\bx}\phi_{k,\ell}&=4\pi\varphi_{k}\overline{\varphi_{\ell}}
  \,,\
  1\le k,\ell\le M\,.
 \end{align}
Since \(V(\bx)={}-\sum_{k=1}^{L}Z_k|\bx-R_k|^{-1}\) is real analytic
on \(\R^3\setminus\{R_1,\ldots,R_L\}\),
\eqref{eq:HF-augmented1}--\eqref{eq:HF-augmented2} shows
that \(\{\varphi_i,\phi_{k,\ell}\}_{i,k,\ell}\) is a solution of an
analytic nonlinear elliptic system of PDE's on
\(\R^3\setminus\{R_1,\ldots,R_L\}\). It follows (from \cite{Morrey,
  MorreyBook} or the method in \cite{Kato}) that
\(\{\varphi_i\}_{i=1,\ldots,M}\) and \(\{\phi_{k,\ell}\}_{1\le
  k<\ell\le M}\) are real analytic in
\(\R^{3}\setminus\{R_1,\ldots,R_L\}\). This is the standard proof that
solutions to the multiconfiguration equations
\eqref{eq:MC-SCF1}--\eqref{eq:MC-SCF2} are real analytic away from the
origin in \(\R^3\) \cite{Lewin,Friesecke}. 

Recall that \(R_1=0\in\R^3\). Note that
\eqref{eq:HF-augmented1}--\eqref{eq:HF-augmented2},
\eqref{eq:Laplace-formula},  
and \eqref{eq:norms} imply that
\begin{align}
  \label{eq:HF-augmented-KS1}
  \nonumber
  \gamma_i\big({}-\Delta_{\by}+4|\by|^2V_K\big)(\varphi_i)_K
  +\sum_{i,j,k,\ell=1}^{M}A_{i j k \ell}\,
   4|\by|^2(\phi_{k,\ell})_K(\varphi_j)_K
  \\\qquad
  {}-4|\by|^2\sum_{j=1}^{M}\lambda_{i,j}(\varphi_{j})_K=0
  \,,\
  1\le i\le M\,,\\\label{eq:HF-augmented-KS2}
  \qquad\qquad
  {}-\Delta_{\by}(\phi_{k,\ell})_K
  =16\pi|\by|^2(\varphi_{k})_K(\overline{\varphi_{\ell}})_{K} 
  \,,\
  1\le k,\ell\le M\,,
\end{align}
with \(V_K, (\varphi_{i})_{K}\), and \((\phi_{k,\ell})_K\) defined by
\eqref{eq:def-compo}. 

Since the functions involved do not have the sufficient regularity for
\eqref{eq:Laplace-formula} to be applied directly, the above deduction
of \eqref{eq:HF-augmented-KS1}--\eqref{eq:HF-augmented-KS2} is
slightly incomplete. One can make a rigorous proof using 
Lemma~\ref{lem:2} and Remark~\ref{rem:ext-K} in Appendix~\ref{sec:KS}
below. This was carried out in  \cite[pp.~6--7]{KS} in a similar
setting and details are therefore omitted here.

Since (using \eqref{eq:norms})
\begin{align}\label{eq:V_K}
  4|\by|^2V_K(\by)
  ={}-4Z_1-\sum_{k=2}^{L}\frac{4Z_k|\by|^2}{|K(\by)-R_k|}
\end{align}
is real analytic in a neighbourhood of \(0\in\R^3\) (recall
\eqref{eq:norms}),
\eqref{eq:HF-augmented-KS1}--\eqref{eq:HF-augmented-KS2} shows that  
\begin{align}\label{eq:fct_K}
  \{(\varphi_i)_{K},
  (\phi_{k,\ell})_{K}\}_{1\le i,k,\ell\le M}
\end{align} is a solution of an  
analytic nonlinear elliptic system of PDE's on
some ball \(B_4(0,R)\subset\R^4\). As before, it follows that
\begin{align}\label{eq:again}
  \{(\varphi_i)_{K}\}_{1\le i\le M} \quad \text{ and }\quad
  \{(\phi_{k,\ell})_{K}\}_{1\le k, \ell\le M}
\end{align}
are real analytic in
\(B_4(0,R)\subset \R^4\). 
Proposition~\ref{prop:one particle} below, proved in \cite{KS}, then
implies the statement of Theorem~\ref{thm:main}. This finishes the
proof of the theorem. 
\end{pf*}
\begin{prop}[\protect{\cite[Proposition~4.1]{KS}}]\label{prop:one
    particle} 
Let $U \subset \R^3$ be open with $0 \in U$, and let $\varphi:U\to \C$
be a function. Let ${\mathcal{U}} = K^{-1}(U) \subset \R^4$, with
\(K:\R^4\to\R^3\) from \eqref{eq:KSbis}, and suppose that
\begin{align}\label{assum:analy}
  \varphi_K = \varphi \circ K:\mathcal{U}\to \C
\end{align}
is real analytic. 

Then there exist functions $\varphi^{(1)}, \varphi^{(2)}$, real
analytic in a neighbourhood of $0 \in \R^3$, such that 
\begin{align}\label{res:prop}
  \varphi(\bx) = \varphi^{(1)}(\bx)+ |\bx|\varphi^{(2)}(\bx)\,.
\end{align}
\end{prop}

\appendix
\section{The Kustaanheimo-Stiefel transform}
\label{sec:KS}
The KS--transform turns out to be a very useful and natural tool for
the investigation of Schr\"odinger equations with Coulombic
interactions (we refer to \cite{KS} for references on this).
In particular \eqref{eq:norms} and the following lemma
are important for our proofs.  Most of the facts stated here are
well-known (see e.g.~\cite[Appendix A]{HelfferEtAl}).
\begin{lemma}[{\cite[Lemma~3.1]{KS}}]\label{lem:2}
  Let \(K:\R^4\to\R^3\) be defined as in \eqref{eq:KSbis}, let
  \(f:\R^3\to\C\) be any \(C^2\)-function, and define
  \(f_K:\R^4\to\C\) by \eqref{eq:def-compo}.
 
  \noindent{\rm (a)} Then \eqref{eq:Laplace-formula} holds:
\begin{align} \label{eq:Laplace-formulaBIS}
  (\Delta f)(K(\by))&=\frac{1}{4|\by|^2}\,\Delta f_K(\by)\,.
\end{align}
\noindent{\rm (b)} 
Furthermore,
let \(U=B_3(0,r)\subset\R^3\) for
\(r\in(0,\infty]\). Then, for \(\phi\in C_0(\R^3)\) (continuous with
compact support),
\begin{align}
  \label{eq:L-2-formula}
  \int_{K^{-1}(U)}|\phi(K(\by))|^2\,d\by =
  \frac{\pi}{4}\int_{U}\frac{|\phi(\bx)|^2}{|\bx|}\,d\bx\,. 
\end{align}
In particular, 
\begin{align}
  \label{eq:L-2-abs}
  \big\||\by|\phi_K\big\|^2_{L^2(K^{-1}(U))}=\frac{\pi}{4}\|\phi\|^2_{L^2(U)}\,.
\end{align}
\end{lemma}
\begin{remark}[{\cite[Remark~3.2]{KS}}]\label{rem:ext-K}
By a density argument, the isometry \eqref{eq:L-2-abs} allows  to
extend the composition by \(K\) given by \eqref{eq:def-compo} (the
pull-back \(K^*\) by \(K\)) to a map
\begin{align*}
  K^{*}: L^2(U, d\bx)&\to L^2(K^{-1}(U), \tfrac{4}{\pi}|\by|^2d\by)\\
  \phi&\mapsto \phi_K
\end{align*}
in the case when \(U=B_3(0,r), r\in(0,\infty]\). 
This makes \(\phi_K\) well-defined for any \(\phi\in L^2(U)\). 
Furthermore, if \(\phi_n\to\phi\) in
\(L^2(U)\), then, for all \(g\in C^{\infty}(K^{-1}(U))\) (\(g\in
C_{0}^{\infty}(K^{-1}(U))\), if \(r=\infty\)),
\begin{align}
  \label{eq:test-conv}
  \lim_{n\to\infty}\int_{K^{-1}(U)} g(\by)(\phi_n)_K(\by)\,d\by
  =\int_{K^{-1}(U)} g(\by)\phi_K(\by)\,d\by\,.
\end{align}
This follows from Schwarz' inequality and \eqref{eq:L-2-abs},
\begin{align*}
  \Big|\int_{K^{-1}(U)}&g(\by)\big((\phi_n)_K(\by)-\phi_K(\by)\big)\,d\by\Big|
  \\&\le
   \Big(\int_{K^{-1}(U)}\frac{|g(\by)|^2}{|\by|^2}\,d\by\Big)^{1/2}
   \big\||\by|\big((\phi_n)_K-\phi_K\big)\big\|_{L^2(K^{-1}(U))}
  \\&=
  \frac{\sqrt{\pi}}{2}\Big(\int_{K^{-1}(U)}\frac{|g(\by)|^2}{|\by|^2}\,d\by\Big)^{1/2} 
  \|\phi_n-\phi\|_{L^2(U)}\to0\,,\ n\to\infty\,.
\end{align*}
Here the \(\by\)-integral clearly converges since \(g\in
C^{\infty}(\R^4)\) (\(g\in
C_{0}^{\infty}(\R^4)\), if \(r=\infty\)).
\end{remark}

\begin{acknowledgement}
This research was (partially) completed while T{\O}S was visiting the
Institute for Mathematical Sciences, National University of Singapore
in 2008. SF is supported by the Danish 
Research Council, the Lundbeck Foundation and by the European Research Council under the 
European Community's Seventh Framework Programme (FP7/2007-2013)/ERC grant agreement 
n$^{\circ}$ 202859. T{\O}S is
partially supported by The 
Danish Natural Science Research
Council, under the grant `Mathematical Physics and Partial Differential
Equations'.
\end{acknowledgement}

\providecommand{\bysame}{\leavevmode\hbox to3em{\hrulefill}\thinspace}
\providecommand{\MR}{\relax\ifhmode\unskip\space\fi MR }
\providecommand{\MRhref}[2]{%
  \href{http://www.ams.org/mathscinet-getitem?mr=#1}{#2}
}
\providecommand{\href}[2]{#2}


\begin{thebibliography}{10}

\bibitem{CancesLeBris}
Eric Canc\`es and Claude Le~Bris, \emph{On the convergence of {SCF} algorithms
  for the {H}artree--{F}ock equations}, M2AN Math. Model. Numer. Anal.
  \textbf{34} (2000), no.~4, 749--774.

\bibitem{Evans}
Lawrence~C. Evans, \emph{Partial {D}ifferential {E}quations}, Graduate Studies
  in Mathematics, vol.~19, American Mathematical Society, Providence, RI, 1998.

\bibitem{Flad}
Heinz-J{\"u}rgen Flad, Wolfgang Hackbusch, and Reinhold Schneider, \emph{Best
  {$N$}-term approximation in electronic structure calculations. {I}.
  {O}ne-electron reduced density matrix}, M2AN Math. Model. Numer. Anal.
  \textbf{40} (2006), no.~1, 49--61.

\bibitem{Flad2}
Heinz-J{\"u}rgen Flad, Reinhold Schneider, and Bert-Wolfgang Schulze,
  \emph{Asymptotic regularity of solutions to {H}artree--{F}ock equations with
  {C}oulomb potentials}, Math. Meth. Appl. Sci. (2008), published online,
  DOI:10.1002/mma.1021.

\bibitem{Fock}
Vladimir~A. Fock, \emph{N{\"a}hrungsmethode zur {L}{\"o}sung des
  quantenmechanischen {M}ehrk{\"o}rperproblems}, Z. Phys. \textbf{61} (1930),
  126--148.

\bibitem{non-iso}
S{\o}ren Fournais, Maria Hoffmann-Ostenhof, Thomas Hoffmann-Ostenhof, and
  Thomas {\O}stergaard~S{\o}rensen, \emph{Non-{I}sotropic {C}usp {C}onditions
  and {R}egularity of the {E}lectron {D}ensity of {M}olecules at the {N}uclei},
  Ann. Henri Poincar\'e \textbf{8} (2007), no.~4, 731--748.

\bibitem{analytic}
S{\o}ren Fournais, Maria Hoffmann-Ostenhof, Thomas Hoffmann-Ostenhof, and
  Thomas~{\O}stergaard S{\o}rensen, \emph{Analyticity of the density of
  electronic wavefunctions}, Ark. Mat. \textbf{42} (2004), no.~1, 87--106.

\bibitem{Monster}
\bysame, \emph{Sharp {R}egularity {R}esults for {C}oulombic {M}any-{E}lectron
  {W}ave {F}unctions}, Commun. Math. Phys. \textbf{255} (2005), no.~1,
  183--227.

\bibitem{KS}
\bysame, \emph{Analytic {S}tructure of {M}any-{B}ody {C}oulombic {W}ave
  {F}unctions}, Commun. Math. Phys. (accepted for publication, 2008), (preprint
  {\tt arXiv:0806.1004v1}).

\bibitem{thirdder}
S{\o}ren Fournais, Maria Hoffmann-Ostenhof, and Thomas
  {\O}stergaard~S{\o}rensen, \emph{Third {D}erivative of the {O}ne-{E}lectron
  {D}ensity at the {N}ucleus}, Ann. Henri Poincar\'e \textbf{9} (2008), no.~7,
  1387--1412.

\bibitem{Friesecke2}
Gero Friesecke, \emph{The {M}ulticonfiguration {E}quations for {A}toms and
  {M}olecules: {C}harge {Q}uantization and {E}xistence of {S}olutions}, Arch.
  Ration. Mech. Anal. \textbf{169} (2003), no.~1, 35--71.

\bibitem{Friesecke}
\bysame, \emph{On the infinitude of non-zero eigenvalues of the single-electron
  density matrix for atoms and molecules}, R. Soc. Lond. Proc. Ser. A Math.
  Phys. Eng. Sci. \textbf{459} (2003), no.~2029, 47--52.

\bibitem{Hartree}
Douglas~R. Hartree, \emph{The wave mechanics of an atom with a non-{C}oulomb
  central field. {P}art {I}. {T}heory and methods}, Proc. Camb. Phil. Soc.
  \textbf{24} (1928), 89--132.

\bibitem{HelfferEtAl}
Bernard Helffer, Andreas Knauf, Heinz Siedentop, and Rudi Weikard, \emph{On the
  absence of a first order correction for the number of bound states of a
  {S}chr\"odinger operator with {C}oulomb singularity}, Comm. Partial
  Differential Equations \textbf{17} (1992), no.~3-4, 615--639.

\bibitem{Kato}
Keiichi Kato, \emph{New idea for proof of analyticity of solutions to analytic
  nonlinear elliptic equations}, SUT J. Math. \textbf{32} (1996), no.~2,
  157--161.

\bibitem{KatoBook}
Tosio Kato, \emph{Perturbation theory for linear operators}, Classics in
  Mathematics, Springer-Verlag, Berlin, 1995, Reprint of the 1980 edition.

\bibitem{LeBris}
Claude Le~Bris, \emph{A general approach for multiconfiguration methods in
  quantum molecular chemistry}, Ann. Inst. H. Poincar\'e Anal. Non Lin\'eaire
  \textbf{11} (1994), no.~4, 441--484.

\bibitem{LeBris2}
\bysame, \emph{Computational chemistry from the perspective of numerical
  analysis}, Acta Numer. \textbf{14} (2005), 363--444.

\bibitem{LeBrisLions}
Claude Le~Bris and Pierre-Louis Lions, \emph{From atoms to crystals: a
  mathematical journey}, Bull. Amer. Math. Soc. (N.S.) \textbf{42} (2005),
  no.~3, 291--363 (electronic).

\bibitem{Lewin}
Mathieu Lewin, \emph{Solutions of the {M}ulticonfiguration {E}quations in
  {Q}uantum {C}hemistry}, Arch. Ration. Mech. Anal. \textbf{171} (2004), no.~1,
  83--114, (preprint {\tt mp-arc:~02-243}).

\bibitem{Lieb84a}
Elliott~H. Lieb, \emph{Atomic and {M}olecular {N}egative {I}ons}, Phys. Rev.
  Lett. \textbf{52} (1984), no.~5, 315--317.

\bibitem{Lieb84b}
\bysame, \emph{Bound on the maximum negative ionization of atoms and
  molecules}, Phys. Rev. A \textbf{29} (1984), no.~6, 3018--3028.

\bibitem{LiebSimonHF}
Elliott~H. Lieb and Barry Simon, \emph{The {H}artree-{F}ock {T}heory for
  {C}oulomb {S}ystems}, Commun. Math. Phys. \textbf{53} (1977), no.~3,
  185--194.

\bibitem{Lions}
Pierre-Louis Lions, \emph{Solutions of {H}artree-{F}ock {E}quations for
  {C}oulomb {S}ystems}, Commun. Math. Phys. \textbf{109} (1987), no.~1, 33--97.

\bibitem{Morrey}
Charles~B. Morrey, Jr., \emph{On the analyticity of the solutions of analytic
  non-linear elliptic systems of partial differential equations. {I}.
  {A}nalyticity in the interior.}, Amer. J. Math. \textbf{80} (1958), 198--218.

\bibitem{MorreyBook}
\bysame, \emph{Multiple {I}ntegrals in the {C}alculus of {V}ariations}, Die
  Grundlehren der mathematischen Wissenschaften, Band 130, Springer-Verlag New
  York, Inc., New York, 1966.

\bibitem{RS2}
Michael Reed and Barry Simon, \emph{Methods of modern mathematical physics.
  {II}. {F}ourier analysis, self-adjointness}, Academic Press [Harcourt Brace
  Jovanovich Publishers], New York, 1975.

\bibitem{Slater}
John~C. Slater, \emph{A note on {H}artree's method}, Phys. Rev. \textbf{35}
  (1930), 210--211.

\bibitem{Zhislin}
Grigorii~M. {\v{Z}}islin, \emph{A study of the spectrum of the {S}chr\"odinger
  operator for a system of several particles ({R}ussian)}, Trudy Moskov. Mat.
  Ob\v s\v c. \textbf{9} (1960), 81--120.

\end{thebibliography}
\end{document}